\documentstyle[aps,epsf]{revtex}
\begin{document}
\twocolumn[\hsize\textwidth\columnwidth\hsize\csname@twocolumnfalse\endcsname
\title{Sliding Phases in XY-Models, Crystals, and Cationic Lipid-DNA
Complexes}
\author{C. S. O'Hern*, T. C. Lubensky*, and J. Toner$^+$}
\address{{\rm *}~Department of Physics and Astronomy,
University of Pennsylvania, Philadelphia PA  19104-6396}
\address{$^+$~Institute for Theoretical Science, Materials Science
Institute, and Department of Physics, \\
University of Oregon, Eugene OR
97403}
\date{\today}
\maketitle

\begin{abstract}
We predict the existence of a totally new class of phases in weakly-coupled,
three-dimensional stacks of two-dimensional ($2$D) $XY$-models.  These
``sliding phases'' behave essentially like decoupled, independent
$2$D $XY$-models with precisely zero free energy cost associated with
rotating spins in one layer relative to those in neighboring layers.
As a result, the two-point spin correlation function decays algebraically
with in-plane separation.  Our results, which contradict past studies
because we include higher-gradient couplings between layers, also
apply to crystals and may explain recently observed behavior in cationic
lipid-DNA complexes.
\\ {\sl Pacs: 61.30.Cz,61.30.Jf,64.70.Md}
\end{abstract}
\vskip2pc]

\par

Spatial dimensionality greatly affects the nature
of order in condensed matter systems. Three-dimensional ($3$D)
$XY$-systems such as superfluids and ferromagnets have true long-range
order with divergent correlation lengths at a second-order transition
separating the high-temperature disordered phase from the
low-temperature ordered phase.  Two-dimensional ($2$D) $XY$-systems,
in contrast, exhibit power-law decay of correlations in the
low-temperature phase.  At high temperatures beyond the
Kosterlitz-Thouless (KT) transition temperature, thermally excited
vortices destroy the quasi-long-range order and cause correlations to
decay exponentially\cite{KT,nelson}.

Many experimentally realizable systems such as layered
superconductors\cite{superconductors}, free-standing liquid-crystal
films\cite{pershan}, and lyotropic smectics with internal membrane
order can be viewed as stacks of two-dimensional layers with
interlayer couplings \cite{koltover} that can be varied substantially,
for example by changing the layer spacing.  What is the phase behavior
of such a system?  If there is no coupling between layers, each will
exhibit $2$D behavior; if the coupling is strong, the system will
exhibit $3$D behavior.  Shortly after the discovery of
the KT transition, it was suggested that a weakly-coupled stack of
$XY$-models might behave over some temperature range as a stack of
decoupled layers, i.e., that such a system could proceed from
three-dimensional behavior at low temperatures to a $2$D power-law
phase at intermediate temperatures to a disordered phase at high
temperatures\cite{superconductors,layeredspin}.

Subsequent studies, however, demonstrated that if the only interlayer
couplings are Josephson (i.e., proportional to $\cos (\theta_n -
\theta_{n+1} )$, where $\theta_n$ is the $XY$-angle variable in layer
$n$), the intermediate $2$D power-law phase is squeezed out, and the
system goes directly from the $3$D long-range ordered phase to the
disordered one with increasing temperature.  This happens because the
``decoupling temperature'' $T_d$ above which the Josephson coupling
becomes irrelevant is greater than the Kosterlitz-Thouless temperature
$T_{KT}$ below which the $2$D ordered phase is stable against vortex
unbinding.  Thus, the temperature window $T_d < T < T_{KT}$ over which
the $2$D sliding phase can exist disappears.

In this paper, we revisit this old and seemingly dead issue and show
that a thermodynamically stable phase exhibiting $2$D power-law
correlations is in fact possible. The new ingredient in our analysis,
which was not present in previous treatments, is competing
higher-order gradient couplings between layers\cite{twolayer}.  These
gradient couplings, in the absence of Josephson couplings between
layers, produce two-point correlation functions that are identical in
form to those of a stack of decoupled $2$D layers.  We will refer to
this phase as a {\it sliding} and not a decoupled phase because the
$XY$-angle variables in different layers can slide relative to each
other without changing the energy of the system and because nonzero
couplings between layers (though not of the Josephson type) are, in
fact, present in our model, and furthermore, necessary for the
existence of this phase.  Remarkably, it is possible through judicious
tuning of interlayer gradient couplings to satisfy $T_d <T_{KT}$ and
produce a stable sliding phase for $T_d < T < T_{KT}$.

This investigation into whether or not a sliding phase of $XY$-models
exists was inspired by recent work on the possible sliding columnar
phase in cationic lipid-DNA complexes\cite{SC}.  In these complexes,
DNA molecules are intercalated between lipid bilayers and, within each
layer, the molecules are situated on a one-dimensional lattice.
Experiments may be consistent with the existence of a sliding
columnar phase in which lattices in neighboring layers are able to
slide over each other without energy cost\cite{experiment} in
complete analogy to the sliding phase just described for $XY$-models.
Indeed, we have shown theoretically, by methods analogous to those
presented here, that it is possible to have a sliding columnar phase
in these complexes and, in addition, possible to have a sliding phase
in a layered crystal.  Details on these two systems will be presented
in a future publication\cite{hexagonal}; for the remainder of this
paper, we will focus on $XY$-models.

The traditional theory for a stack of $XY$-models begins with
sum of independent $XY$-Hamiltonians
\begin{equation}
\label{HXY}
{\cal H}_0 = {K \over 2} \sum_n \int d^2 r
\left[{\mbox{\boldmath{$\nabla$}}}_{\perp} \theta_n({\bf r})
\right]^2,
\end{equation}
where ${\bf r} = (x,y,0)$ is a point in the $x$-$y$ plane and
${\mbox{\boldmath{$\nabla$}}}_{\perp}$ is the gradient operator acting
on these two coordinates.  Josephson-like couplings between layers are
then added; these are given by
\begin{equation}
\label{Josephson}
{\cal H}_{J}[s_n] = - V_J[s_n] \int d^2r \cos\left[
\sum_p s_p \theta_{n+p}
({\bf r})\right],
\end{equation}
where $s_n$ is an integer-valued function of layer number $n$ satisfying
$\sum_n s_n =0$ if there are no external fields inducing long-range order.
If all $V_J[s_n]$ are zero, then in the low-temperature
phase $\langle \theta_n^2({\bf r})\rangle_0 =\eta \log(L/b)$
and $\cos[\theta_n ( {\bf r} ) - \theta_n ( 0)] \sim r^{-\eta}$, where
\begin{equation}
\label{eta}
\eta = {T \over 2\pi K},
\end{equation}
$L$ is the sample width, $b$ is a short-distance cutoff in the
$x$-$y$ plane, and $\langle . \rangle_0$ refers to an average with
respect to ${\cal H}_0$. The averages of the Josephson Hamiltonians with
respect to ${\cal H}_0$ scale as $\langle {\cal H}_J[s_n]\rangle \sim L^{2 -
\eta[s_n]}$ where $\eta [s_n] = {\eta \over 2} \sum_p s_p^2$.  Clearly, the
most relevant Josephson coupling is the one with the smallest value
of $\eta [s_n]$, which results when $s_n$ is non-zero on the smallest number
of planes.  Since $\sum_n s_n =0$, the smallest value of $\eta [s_n]$ is
obtained for couplings between two layers separated by $p$ layers
with $s_n = s_n^p =\delta_{n,0} -\delta_{n,p}$.  For these two-layer
couplings, $\eta^p = \eta[s_n^p] = \eta$ for all values of $p$.  Thus,
the decoupling temperature above which all Josephson couplings are
irrelevant is $T_d = 4 \pi K$.  The $2D$ KT transition for decoupled
layers is $T_{KT} = \pi K/2$; this implies $T_{KT} / T_d = 1/8 <1$,
and there is no decoupled phase with power-law correlations.

Josephson couplings are not, however, the only ones permitted
by symmetry. Gradients of $\theta_n$ in different layers may also be coupled.
The Hamiltonian for the ideal sliding
phase is ${\cal H}_S = {\cal H}_0 + {\cal H}_g$, where
\begin{equation}
\label{HXYint}
{\cal H}_g = {1 \over 2}
\sum_{n,m} \int d^2r~{U_m \over 2}
\left[{\mbox{\boldmath{$\nabla$}}}_{\perp}
\left( \theta_{n+m} ({\bf r}) -
\theta_{n}({\bf r})\right)\right]^2.
\end{equation}
This Hamiltonian is invariant with respect to $\theta_n({\bf
r}) \rightarrow \theta_n({\bf r}) + \psi_n$ for any constant $\psi_n$, i.e.,
the energy is unchanged when angles in different layers slide relative
to one another by arbitrary amounts.  The sliding Hamiltonian
can be written  as
\begin{equation}
\label{sliding}
{\cal H}_S  = {1 \over 2} \sum_{n n'}\int d^2 r~K_{n n'}
{\mbox{\boldmath{$\nabla$}}}_{\perp} \theta_n({\bf r}) \cdot
{\mbox{\boldmath{$\nabla$}}}_{\perp} \theta_{n'}({\bf r}), \\
\end{equation}
where $K_{n n'} = K f_{n - n'}$ with $f_n = (1 + \sum_m \gamma_m)
\delta_{n,0} - \case{1}{2} \sum_m \gamma_m (\delta_{n,m} + \delta_{n,-m})$
and $\gamma_m = U_m/K$. Also, the Fourier transform
\begin{equation}
\label{Kqz}
f(k) = 1 + \sum_m \gamma_m (1 - \cos km )
\end{equation}
of the reduced coupling $f_n$ will be used extensively below.

Correlations in the sliding phase can easily be calculated from
Eq.~(\ref{sliding}). We find
\begin{equation}
\langle \theta_n ({\bf r}) \theta_{n'} ( {\bf r}' ) \rangle_S
=\eta f^{-1}_{n - n'} [ \ln (L/b) - E(|{\bf r} - {\bf r}'|) ] ,
\label{thetacorr}
\end{equation}
where $\langle \cdot \rangle_S$ is an average with respect to ${\cal
H}_S$ and $E(r) = \int dq [ 1 - J_0 (q r)]/q$ tends to zero as $r
\rightarrow 0$ and to $\ln (r/b')$ with $b'/b \approx 0.2$ as $ r
\rightarrow \infty$.  The inverse coupling $f_p^{-1}$ is defined by
\begin{equation}
f_p^{-1} = {1 \over \pi} \int_0^{\pi} dk {\cos kp \over f(k)} .
\label{fp}
\end{equation}
Thus we find $\langle \theta_n^2({\bf r}) \rangle_S =
\eta_S(0) \ln(L/b)$ and
\begin{eqnarray}
\label{correlation}
g_S({\bf r},p) & \equiv & \langle [\theta_{n+p}({\bf r}) -
\theta_n(0)]^2\rangle_S \nonumber \\
& = & 2 [ \widetilde{\eta}_S(p) \ln (L/b) +
\eta_S(p) \ln (r/b')]
\end{eqnarray}
for large $r$.  The coefficients of the logarithms are 
\begin{equation}
\widetilde{\eta}_S(p) = \eta (f_0^{-1} - f_p^{-1})~~{\rm and}~~\eta_S(p) =
\eta f_p^{-1}.
\end{equation}
Note that $\eta_S(p) = \eta \delta_{p,0}$ and ${\widetilde \eta}_S(p)
= \eta (1-\delta_{p,0})$ when ${\cal H}_g=0$.  Using
Eq.~(\ref{correlation}) we find that the correlation
function $G_S({\bf r},p) \equiv \left\langle \cos[\theta_{n+p}({\bf r}) -
\theta_n(0)] \right\rangle_S$ satisfies
\begin{eqnarray}
\label{spincorr}
G_S({\bf r},p) & \sim & \cases{(L/b)^{-{\widetilde \eta}(p)}  & $p \ne 0$\cr
       \left(r/b'\right)^{-\eta_S(0)} & $p=0$.\cr}
\end{eqnarray}
Thus, the two-point spin correlation function for spins in different
layers vanishes in the $L \rightarrow \infty$ limit, whereas that for
spins in the same layer has exactly the same form as it would for a
stack of decoupled layers.  Now, however, the exponent $\eta_S(0)$
depends on the detailed form of the interlayer gradient couplings via
$f(k)$.  The two-point spin correlation function is zero for spins in
different layers; nonetheless, nonvanishing couplings between layers
cause other correlation functions that are zero for the totally
decoupled layers to become nonzero.

Having established that the sliding phase (if it has not melted)
behaves like a stack of decoupled layers, we now ask what happens when
the Josephson interlayer couplings of Eq.\ (\ref{Josephson}) are
turned on. From Eq.\ (\ref{thetacorr}), we find that $\langle {\cal
H}_J[s_n]\rangle_S \sim L^{2-{\tilde
\eta}_S [s_n]}$ where ${\widetilde \eta}_S[s_n] =
\eta \sum_{n,n'} s_n s_{n'} f^{-1}_{n - n'}$.
As for the decoupled case, the minimum value of
${\widetilde \eta}_S [s_n]$ is obtained when $s_n= s_n^p$.
Thus the decoupling temperature for couplings with $s_n = s_n^p$ is
\begin{equation}
T_d (p) = {4 \pi K \over f_0^{-1} - f_p^{-1}}
\label{Td}
\end{equation}
which depends on $p$. The temperature above which all Josephson
couplings are irrelevant is $T_d = \max_p T_d ( p )$. We will show later
that this
maximum over all $p$ is finite.

To prove the stability of the sliding phase, we must show that, for
some range of the couplings $U_m$, the decoupling temperature $T_d$
calculated above is less than $T_{KT}$, the temperature at
which vortices unbind. To calculate $T_{KT}$, we must
calculate the vortex energy. This calculation in our model is similar
to that for decoupled layers.  Vortex excitations in individual layers
remain well defined when the layers are coupled, although, when
couplings are sufficiently strong the system becomes truly three
dimensional, and vortices should be viewed as segments of closed
vortex loops.  Defining ${\bf v}_n({\bf r}) =
{\mbox{\boldmath{$\nabla$}}}_{\perp}
\theta_n({\bf r})$, we have
\begin{equation}
\label{lineint}
\oint_{\Gamma} {\bf v}_n \cdot d {\mbox{\boldmath{$\ell$}}} = 2 \pi 
\sum_l k_{n,l},
\end{equation}
where $k_{n,l}$ is the integer strength of the $l$th vortex in the
$n$th layer and $\Gamma$ is a contour in layer $n$ enclosing the
vortices.  Applying Stokes theorem to Eq.~(\ref{lineint}) then gives
\begin{equation}
\label{constraint}
{\mbox{\boldmath{$\nabla$}}}_{\perp} \times {\bf v}_n = m^z_n({\bf r})
{\hat z},
\end{equation}
where
\begin{eqnarray}
\label{m def}
m^z_n({\bf r}) = 2 \pi \sum_{l} k_{n,l}
\delta^2({\bf r}-{\bf r}_{n,l})
\end{eqnarray}
is the vortex density in layer $n$ and ${\bf r}_{n,l}$ gives the
position of each vortex in the $x$-$y$ plane.  We then take the $2$D
curl of both sides of Eq.~(\ref{constraint}) and Fourier transform to
find
\begin{equation}
{\bf v}_n({\bf q}_{\perp}) =  {i\epsilon _{ijz} {q_{\perp j}}
m^z_n({\bf q}_{\perp})  \over q_{\perp}^2}
\end{equation}
Using this result in Eq.~(\ref{sliding}), we obtain the
vortex energy
\begin{eqnarray}
E_V &= & {K \over 2} \sum_{n,n'} f_{n-n'}
\int {d^2q_{\perp}\over (2\pi)^2}
{ m^z_n ({\bf q}_{\perp}) m^z_{n'}(-{\bf q}_{\perp}) \over q_{\perp}^2}
\nonumber\\
& = & \pi K \sum_{n,n'} f_{n-n'} \left(\sum_l k_{n,l}\right)\left(\sum_{l'}
k_{n',l'}\right) \ln (L/b) \nonumber \\
&  & - \pi K \sum_{n,l,n',l'} f_{n-n'} k_{n,l} k_{n',l'} 
E(|{\bf r}_{n,l} -{\bf r}_{n',l'} |) .
\label{Hm}
\end{eqnarray}
Since $f_{n-n'}$ is a positive definite matrix, this equation implies
that in the thermodynamic limit, there must be charge neutrality in
each layer, i.e., $\sum_l k_{n,l} = 0$. The interaction between
like-sign vortices in different layers $n$ and $n'$ is attractive if
$f_{n-n'} <0$ and repulsive if $f_{n-n'}>0$.  Since we assume that
individual layers are stable in the absence of couplings between
layers, $f_0>0$ and like-sign vortices within a single
layer repel.

The vortex energy $E_V$ and Boltzmann statistics imply that the number of
times a given configuration of vortices occurs in the system
scales with system size as $L^{2-\eta_{KT}[\sigma_n]}$, where
\begin{equation}
\eta_{KT}[\sigma_n] ={\pi K \over T} \sum_{n,n'} f_{n-n'} \sigma_n
\sigma_{n'},
\label{etasig}
\end{equation}
$\sigma_n \equiv \sum_\ell k_{n,l}$, and the factor of $L^2$
counts the number of places in the $2$D plane the configuration can be
placed.  Clearly, if $\eta[\sigma_n] < 2$, the particular vortex
configuration $\{\sigma_n\}$ will proliferate. The
``Kosterlitz-Thouless unbinding'' for $\{\sigma_n\}$ therefore occurs
at a temperature
\begin{equation}
T_{KT}[\sigma_n] = {\pi K \over 2} \sum_{n,n'} f_{n-n'} \sigma_n \sigma_{n'}.
\label{Tsig}
\end{equation}
If there is only one vortex in layer $0$, then $\sigma_n \equiv
\sigma_n^0 = \delta_{n,0}$ and $T_{KT}[\sigma_n^0] = \pi K f_0/2$.  If
there is a $+1$ vortex in layer zero and a $\pm1$ vortex in layer $p$,
$\sigma_n \equiv \sigma_n^{p\pm} =
\delta_{n,0}\pm \delta_{n,p}$ and $T_{KT}[\sigma_n^{p\pm}] = \pi K
(f_0 \pm f_p)$.  Note that when $f_p$ is nonzero $T_{KT}[\sigma_n^{p\pm}]$ is
not twice $T_{KT}[\sigma_n^0]$.  In fact it is possible for
$T_{KT}[\sigma_n]$ to be less than $T_{KT}[\sigma_n^0]$ for one or more
configurations $\{\sigma_n\}$.  The interactions between layers lead
to composite multi-layer vortices that cost less energy to create than
a single vortex in an individual layer.  Unbinding of bound pairs of
any set of individual-layer or composite vortices will destroy the
rigidity within those layers.  Thus, the transition temperature to the
disordered state is $T_{KT} =
\min_{\{\sigma_n\}} T_{KT}[\sigma_n]$, and the sliding phase
exists provided
\begin{equation}
\beta = {T_{KT}\over T_d} = {\min_{\sigma_n} T_{KT}[\sigma_n]\over
\max_p T_d(p)} >1.
\label{beta}
\end{equation}

\begin{figure}
\centerline{\epsfbox{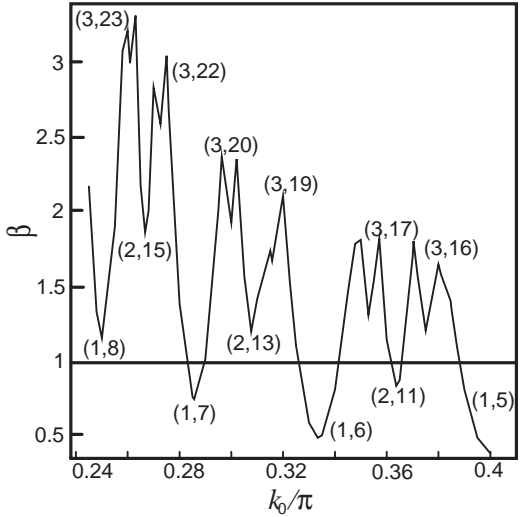}}
\caption{$\beta=T_{KT}/T_d$ is plotted versus $k_0/\pi$.
Local minima near $k_0/\pi = 2l/p$ are labeled
by $(l,p)$.  Other possible integer pairs either do not fall in the
range $0.24 < k_0/\pi < 0.40$ or yield larger values of $\beta$
than those shown above.}
\label{xyfig}
\end{figure}

We will now discuss how the interlayer gradient potentials $U_m$ can
be chosen so that $\beta > 1$. The basic strategy is to choose the
$U_m$ so that $f(k)$ has a minimum near zero at some value of $k$. We
consider a model with both first- and second-neighbor interactions. We
ensure that there is a minimum in $f(k)$ at $k_0$ by requiring
\begin{equation}
f(k_0) = 1 + \gamma_1 \left(1 - \cos k_0\right) +
\gamma_2 \left(1 - \cos 2k_0\right) = \Delta
\label{delta}
\end{equation}
and $f'(k_0) = 0$.  These two conditions determine $\gamma_1$ and
$\gamma_2$ in terms of $k_0$ and $\Delta$.  In the range of $k_0$ and
$\Delta$ we consider, $\gamma_1>0>\gamma_2$ and $\gamma_1 >
|\gamma_2|$.  The minimum can be tuned to zero by taking $\Delta$ to
zero, in which case $f(k_0) = 0$ but $f(k) > 0$ for all $k \ne k_0$.
For small $\Delta$, $f_0^{-1}-f_p^{-1}$ is dominated by values of $k$
near $k_0$, and we have
\begin{eqnarray}
\label{periodic}
f^{-1}_0-f^{-1}_p & \approx & {1-\cos(pk_0)e^{-p \sqrt{\Delta/C}} \over
\sqrt{C \Delta}} \\
& \approx & {p \over C}+{(p k_0 - 2\pi l)^2 \over 2 \sqrt{\Delta C}}, \nonumber
\end{eqnarray}
where the final form is valid for $p k_0 \sim 2\pi l$, $l$ is an
integer, and $C=f''(k_0)/2$.  From Eq.~(\ref{periodic}), we see that
there exists a curve $T_d(p,k_0) \sim [p/C + (p k_0 - 2\pi
l)^2/2\sqrt{\Delta C}]^{-1}$ for each value of $p$ that specifies the
decoupling temperature as a function of $k_0$.  For fixed $k_0$,
$\max_p T_d(p,k_0)$ occurs at $p=[2\pi l/k_0]$ if $0\le \{2\pi
l/k_0\}\le 1/2$ and at $p=[2\pi l/k_0] +1$ if $1/2 <\{2\pi l/k_0\}
<1$, where $[x]$ is the greatest integer less than or equal to $x$ and
$\{x\}=x-[x]$ is the fractional part of $x$.  As a function of $k_0$
near $2\pi l/p$, $T_d(p,k_0)$ reaches a maximum at $k_0=2\pi l/p$ and
decreases sharply away from this point.  Also, in the range of $k_0$
and $\Delta$ we have considered, we can prove that composite like-sign
vortices in nearest-neighbor planes $p$ and $p+1$ with $\sigma_n =
\delta_{n,p} +\delta_{n, p+1}$ are the first to unbind and thus
$T_{KT} = \pi K(f_0 + f_1) = \pi K ( 1 +
\gamma_1/2 + \gamma_2)$.  Since $T_{KT}$ is a smooth function of
$k_0$, we find that $\beta = T_{KT}/T_d$ has sharply-peaked minima at
$k_0 = 2\pi l/p$.  Direct evaluation of $\beta$ for $\Delta = 10^{-5}$
yields $\beta >1$ in the range $0.24 < k_0/\pi < 0.40$ as shown in
Fig.~\ref{xyfig}.

Transitions out of the sliding phase are of the Kosterlitz-Thouless or
roughening type.  The transition to the high-temperature disordered
phase at $T_{KT}$ is controlled by $K$ and the fugacity $y_{++}$ for
composite like-sign vortices in neighboring layers. The transition to
the low-temperature $3D$ ordered phase is controlled by the first
$V_p \equiv V_J [s_n^p]$ to become relevant and by $U_p$.

As we have seen, the Josephson couplings $V_p$ are irrelevant with
respect to the sliding phase for $T_d < T < T_{KT}$.  If all $V_p$ are
set to zero, the two-point correlation function $G_S({\bf r},p)$
vanishes for $p \ne 0$.  Even though the $V_p$ are irrelevant, they
are not zero.  They give rise to nonzero perturbative contributions
to $G_S({\bf r},p)$ even when $p$ is nonzero.  Consider for simplicity
the nearest-neighbor Josephson model ($V_p = V \delta_{p,1}$).  Then
\begin{equation}
\label{expcorr}
G_S({\bf r},p) = \left( V \over 2T\right)^p \int d^2 r_1 \ldots d^2 r_p
e^{-\Phi({\bf r}_1,\ldots,{\bf r}_p, {\bf r})/2},
\end{equation}
where $\Phi({\bf r}_1,\ldots,{\bf r}_p, {\bf r}) = \langle [\Delta
\theta_0(0,{\bf r}_1) + \Delta \theta_1({\bf r}_1, {\bf r}_2)
+\ldots+\Delta \theta_p({\bf r}_p,{\bf r}) ]^2 \rangle_S$ and 
$\Delta \theta_n({\bf r}_1,{\bf r}_2) = \theta_n({\bf r}_1)-
\theta_n({\bf r}_2)$.  Using the fact that the transition at $T_d^+$ is 
a KT transition, it can be shown that Eq.~(\ref{expcorr}) yields an
exponential decay of correlations with $G_S(0,p) \sim e^{-p/\xi_z}$.
The interlayer correlation length $\xi_z$ diverges as $T \rightarrow
T_d^+$ according to $\xi_z \propto 1/(T - T_d)$, i.e. the interlayer
correlation length exponent is $\nu_z = 1$.  This divergence signals
the development of true long-range orientational order in the $3$D
ordered phase below $T_d$.  The derivation of this result will be
presented in a forthcoming publication
\cite{hexagonal}.

The ideas presented here can also be applied to a three-dimensional
stack of two-dimensional crystals\cite{hexagonal}.  An interaction
Hamiltonian analogous to ${\cal H}_g$ in Eq.~(\ref{HXYint}) that
couples gradients of displacements in different layers can
be introduced.  Power-law exponents and dislocation energies again
depend on these couplings, and a sliding crystal phase between
a low-temperature crystalline and a higher-temperature
hexatic phase\cite{birgeneau} is possible.  The sliding crystal phase
is similar to a model once proposed for the smectic B phase in liquid
crystals\cite{degennes}.  Also, interlayer gradient couplings for the
hexatic angle can be introduced to produce a sliding hexatic phase.
Thus the phase sequence $3$D crystal $\rightarrow$ sliding crystal
$\rightarrow$ $3$D hexatic $\rightarrow$ sliding hexatic $\rightarrow$
disordered layers is in principle possible in lamellar systems.

This work was supported in part by the National Science Foundation
under grants DMR97--30405 and DMR--9634596.  J.\ T.\ and T.\ C.\ L.\
thank the Aspen Center for Physics for their Winter Meeting, at which
this work was initiated. We are also grateful to C. Kane for
emphasizing the possibility of melting via composite vortices.

\end{document}